\documentclass{aaMY}
\input epsf.sty
\usepackage{graphicx}
\usepackage{times}

\newcommand{\lsi}    {LS~I+61$^{\circ}$303}
\newcommand{\ltsima} {$\; \buildrel < \over \sim \;$}
\newcommand{\simlt}  {\lower.5ex\hbox{\ltsima}}            
\newcommand{\gtsima} {$\; \buildrel > \over \sim \;$}
\newcommand{\simgt}  {\lower.5ex\hbox{\gtsima}}            

\begin{document}

  \title{ The Be/X-ray Binary LSI+61$^0$303 \\ in terms of Ejector-Propeller Model }

  \author {Radoslav~Zamanov \inst{1}
           \and 
           Josep~Mart\'{\i} \inst{2} 
	   \and 
	   Paolo~Marziani \inst{1} }


  \institute{  Osservatorio Astronomico di Padova, Vicolo dell'Osservatorio 5,
                I-35122 Padova, Italy  \\
		\email{zamanov@pd.astro.it~} ~ ~ \email{marziani@pd.astro.it}
            \and
             Departamento de F\'{\i}sica, EPS,
             Universidad de Ja\'en, C/ Virgen de la Cabeza, 2, E--23071 Ja\'en,
	     Spain \\
             \email{jmarti@ujaen.es}
	     }
\abstract{
We tested the ejector-propeller model of the Be/X-ray binary LSI+61$^0$303 
by using the parameters predicted by the model in the calculations of the X-ray and radio
variability. The results are: (1) in terms of the Ejector-Propeller model, the X-ray maximum is due to 
the periastron passage; (2) the radio outburst can be really a result of the transition from the
propeller to the ejector regimes; (3) the radio outburst will delay with respect to the X-ray maximum every orbital period.
The proposed scenario seems to be in good agreement with the observations.  
}

\date{ Poster - 2$^{nd}$\ CNOC "Astrofisica degli Oggetti compatti" - 
Septmeber 19$-$21, 2001, Bologna, Italy }

\maketitle

\markboth{Zamanov, Mart\'{\i}, Marziani : Be/X-ray binary LS~I+61$^0$303}{}

\section{Flaring behaviour of \lsi\ }

\lsi\ (\object{V615 Cas}, \object{GT0236+620})  is a well known 
Be/X-ray binary with periodic radio outbursts
every 26.5 d, assumed to be the orbital period. 
After the discovery of such periodic events  (Gregory \& Taylor 1978; 
Taylor \& Gregory 1982) and the first model (Maraschi \& Treves, 1981),
this massive system has been studied extensively since nearly two decades ago.
Hereafter, we will use the late radio phase ephemeris 
$P=26.4917\,d$ and phase zero at JD2443366.775 (Gregory, Peracaula \& Taylor, 
1999).

The flaring radio emission of \lsi\ has been modeled by Paredes et al. (1991) 
as synchrotron radiation from an expanding plasmon containing relativistic particles 
and magnetic fields. The dependence of the radio outburst flux density on
the frequency, the time delay in the outburst peak, and the general shape of
the radio light curves are in general well accounted for by continuous injection of
synchrotron emitting electrons into the plasmon volume.
However, the formation details of such plasmon were beyond the scope of the Paredes et al. (1991) model.
In fact, its initial physical properties were considered as the free parameters to be fitted.
According to Zamanov (1995), the genesis of such plasmon can be interpreted
as a result of a transition in the accretion regime of the neutron star as it probes
different parts of the Be envelope in an eccentric orbit. 
The proposed transition would be from so called propeller (P) to the Ejector (E) regime.
In this context, propeller means accretion onto the magnetosphere and ejector 
is often named ``young radio pulsar''.  

Our purpose here is to try to check the agreement between the plasmon 
parameters predicted by the theory of a PE transition and those required to
reproduce the observed radio light curves using the Paredes et al. (1991) model.
We advance here that the agreement seems to be good at least qualitatively, 
including the possibility to explain the X-ray luminosity and variability. 
This agreement provides a consistent physical interpretation 
for the outburst mechanism in the \lsi\ system. 

\section{Simultaneous Radio and X-ray observations} 

\begin{figure}[htb]
 \mbox{}
 \vspace{15.0cm}
 \includegraphics{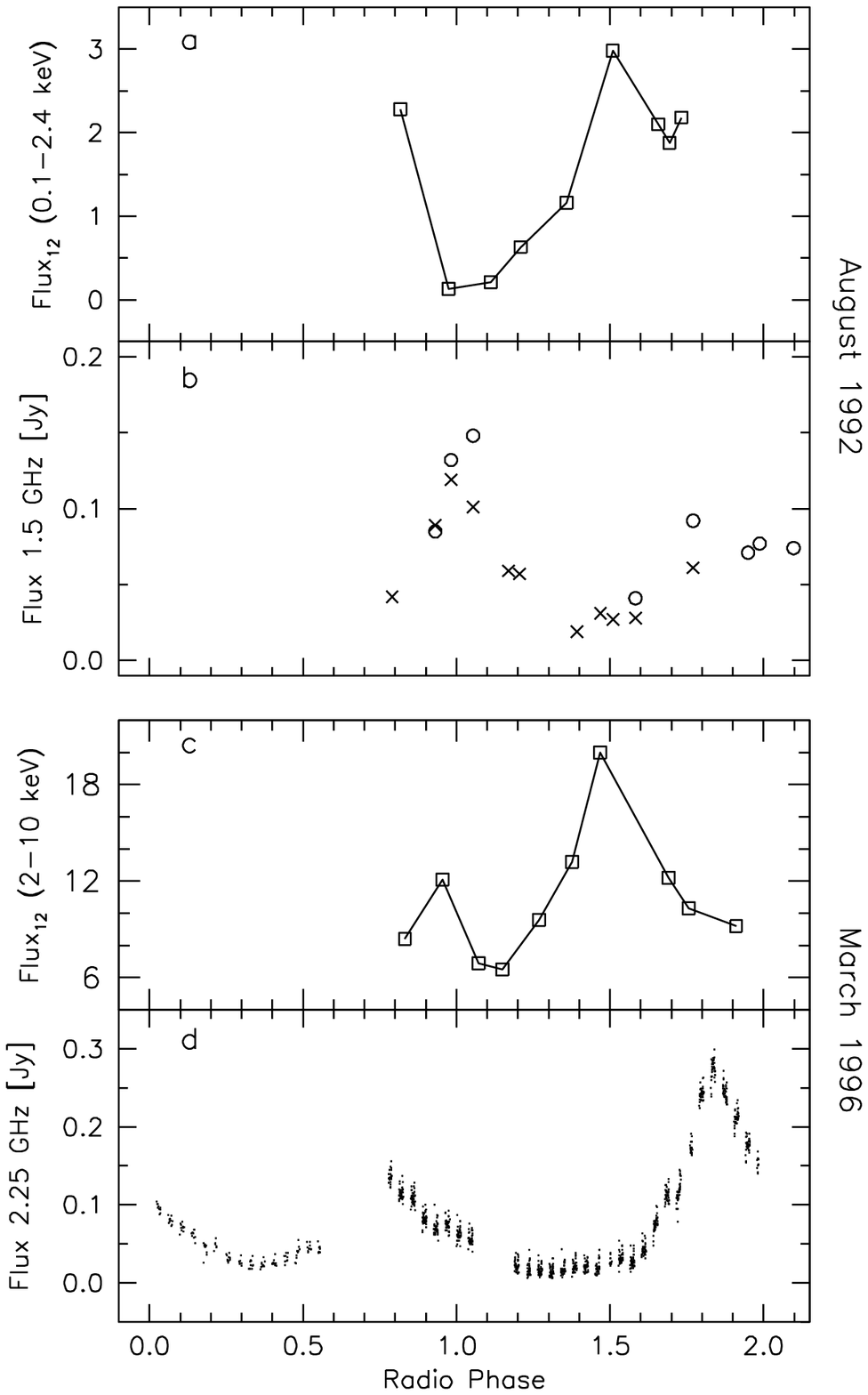}
 \caption[]{ The simultaneous X-ray and radio observations: \\
 {\bf (a)} and  {\bf (b)} -- August 1992, Taylor et al. (1996) \\ 
 {\bf (c)} and  {\bf (d)} -- March 1996, Harrison et al. (2000)\\
 A clear phase shift between X-ray and radio is visible in both cases. \\
 In {\bf (b)} the triangles refer to 1.5 GHz and the crosses  -- to 4.9
 GHz flux density.  
 The X-ray flux is everywhere in units 10$^{-12}$ erg cm$^{-2}$ s$^{-1}$.
 }
 \label{Fiona}
\end{figure}

The variability of the source in the X-ray domain was detected 
in ROSAT  observations (Goldoni \& Mereghetti, 1995). 
On two occasions, the \lsi\  flaring events have been monitored by ground
based radio facilities and space X-ray satellites in a coordinated way.
The results were presented in Taylor et al.(1996) and  Harrison et al.(2000). 
The behavior observed is reproduced in Fig. \ref{Fiona}, using the latest 
ephemeris determination. The observations during 1992 are obtained
with ROSAT (0.5-2 keV) and the VLA (1.5 GHz and 4.9 GHz), 
and those during 1996 - with RXTE (2-10 keV) and 
GBI \footnote{The Green Bank Interferometer is a facility
of the USA National Science Foundation operated by NRAO in support of the
NASA High Energy Astrophysics programs.} (2.25 GHz and 8.3 GHz).
From this figure, it is clear that in both cases
the radio peak is delayed with respect to X-ray maximum. 
 
\section{Mass accretion rate }

\subsection{Bondi-Hoyle accretion}

Usually, to estimate the mass accretion rate onto a compact object 
in wind fed binaries the Bondi-Hoyle-Littleton formula is used in the form:
   \begin{equation}
     \dot M _a=\pi \frac{ (2 G M_{NS})^2 }{V_{rel}^{3}} \rho ,
   \end{equation}
where $M_{NS}$ is the mass of the neutron star, $V_{rel}$ is the NS velocity
relative to the surrounding matter, $\rho$ is the density in the surrounding matter. 

The structure of the Be star disks are not yet well understood, even 
for the most important parameters, such as the velocity and density laws, 
different assumptions have been proposed.  
Following Waters et al. (1988), we will accept a density structure in the form
   \begin{equation}
      \rho= \rho_0 (r/R_*)^{-n} \hskip 0.3cm {\rm and} \hskip 0.3cm   V_r=V_0 (r/R_*)^{n-2},       
   \end{equation}
where $R_*$ is the radius of the Be star and $V_r$ is the outward velocity. 
The density parameter $n$ can take values $n=2\,-\,4$ for different objects.
The initial velocity, $V_0$ (the velocity at $r=R_*$), 
most probably lies between 2 and 20~km~s$^{-1}$. Finally, a Keplerian law is adopted 
concerning the disk rotation:
\begin{equation}
    V_\phi(r) =\sqrt{GM_1/r}       
\end{equation}
where $M_1$ is the mass of the primary component in the system, i.e. the Be star.

\begin{figure*}[htb]
 \mbox{}
 \vspace{12.0cm}
 \includegraphics{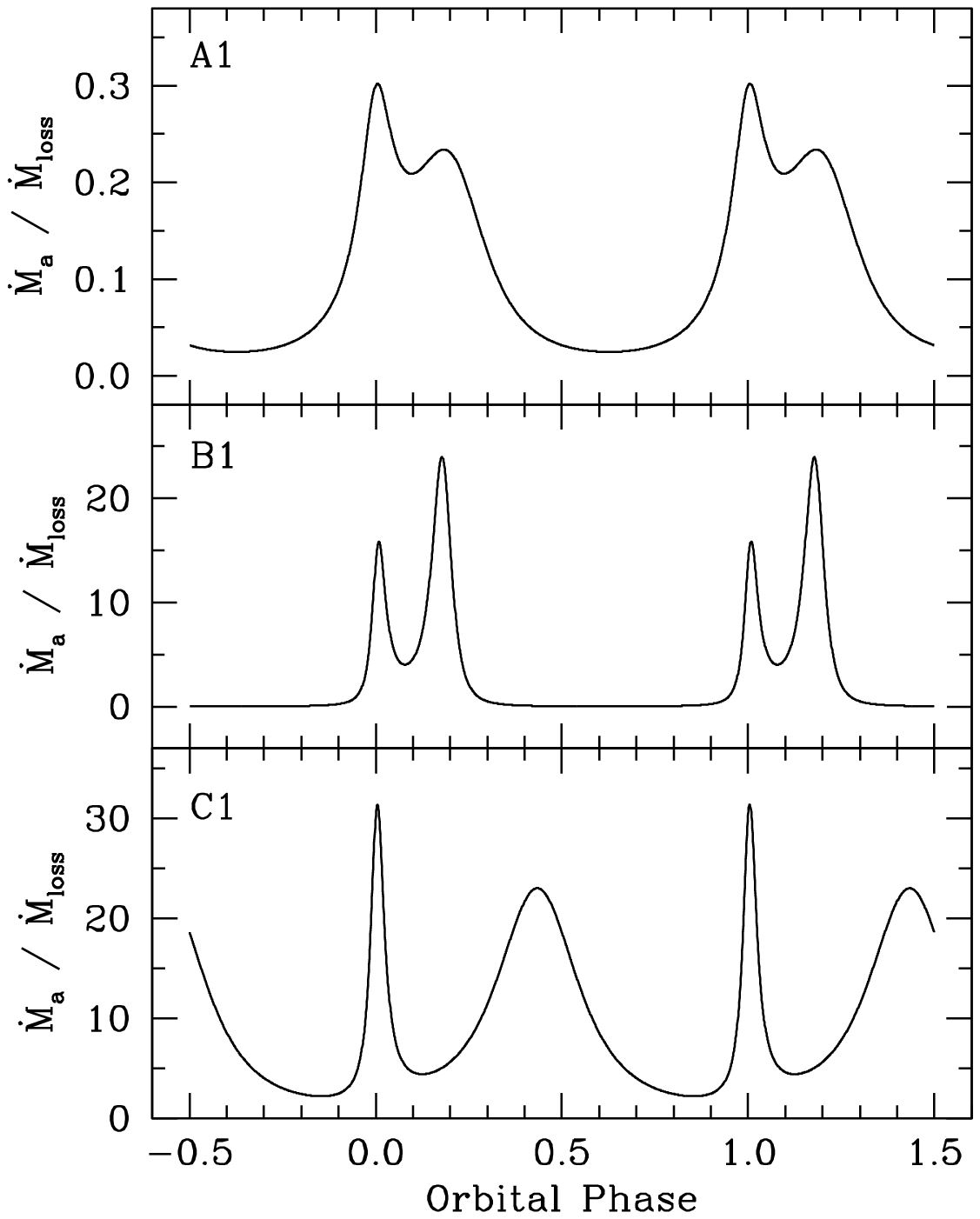}
 \includegraphics{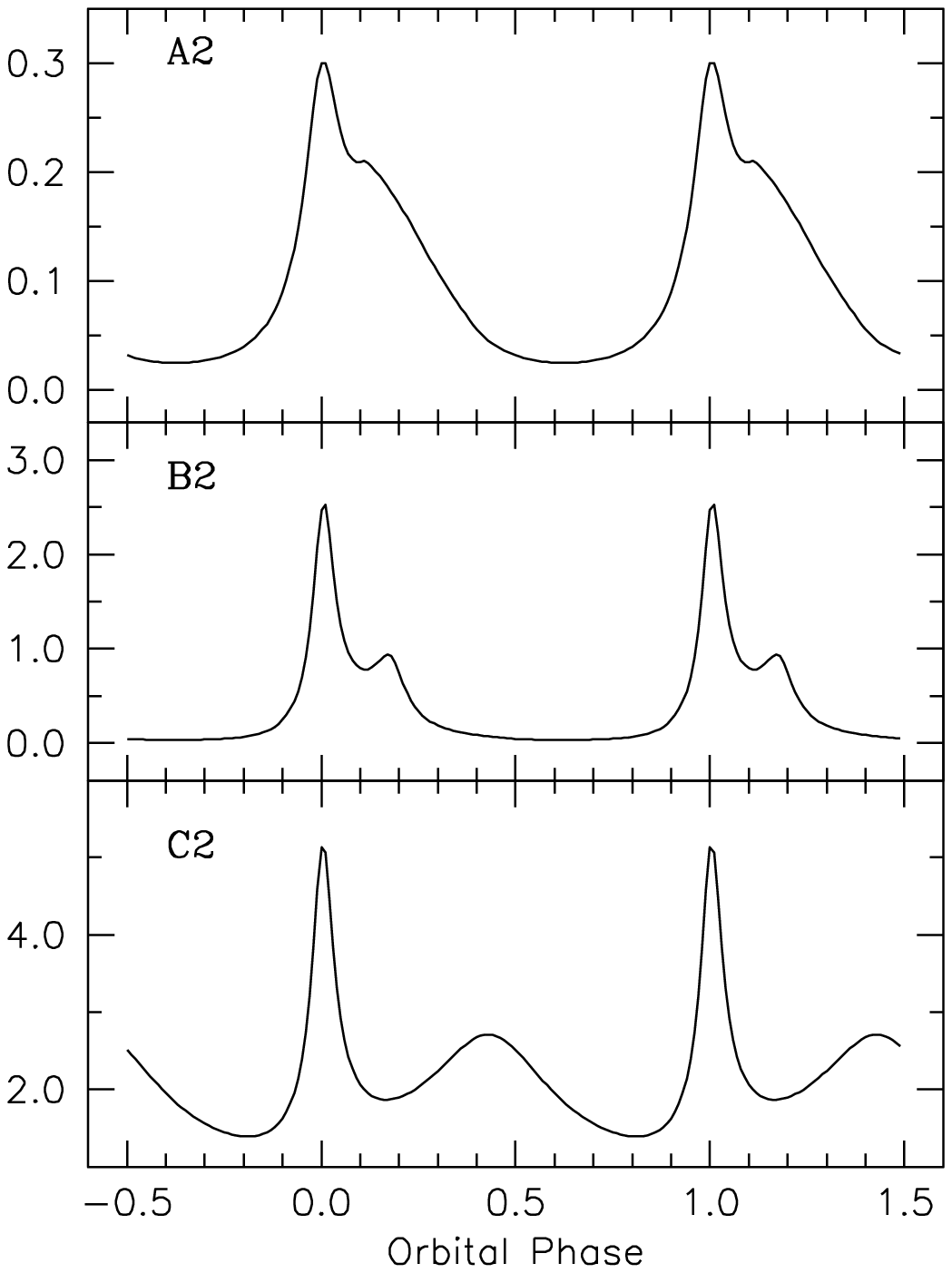}
 \includegraphics{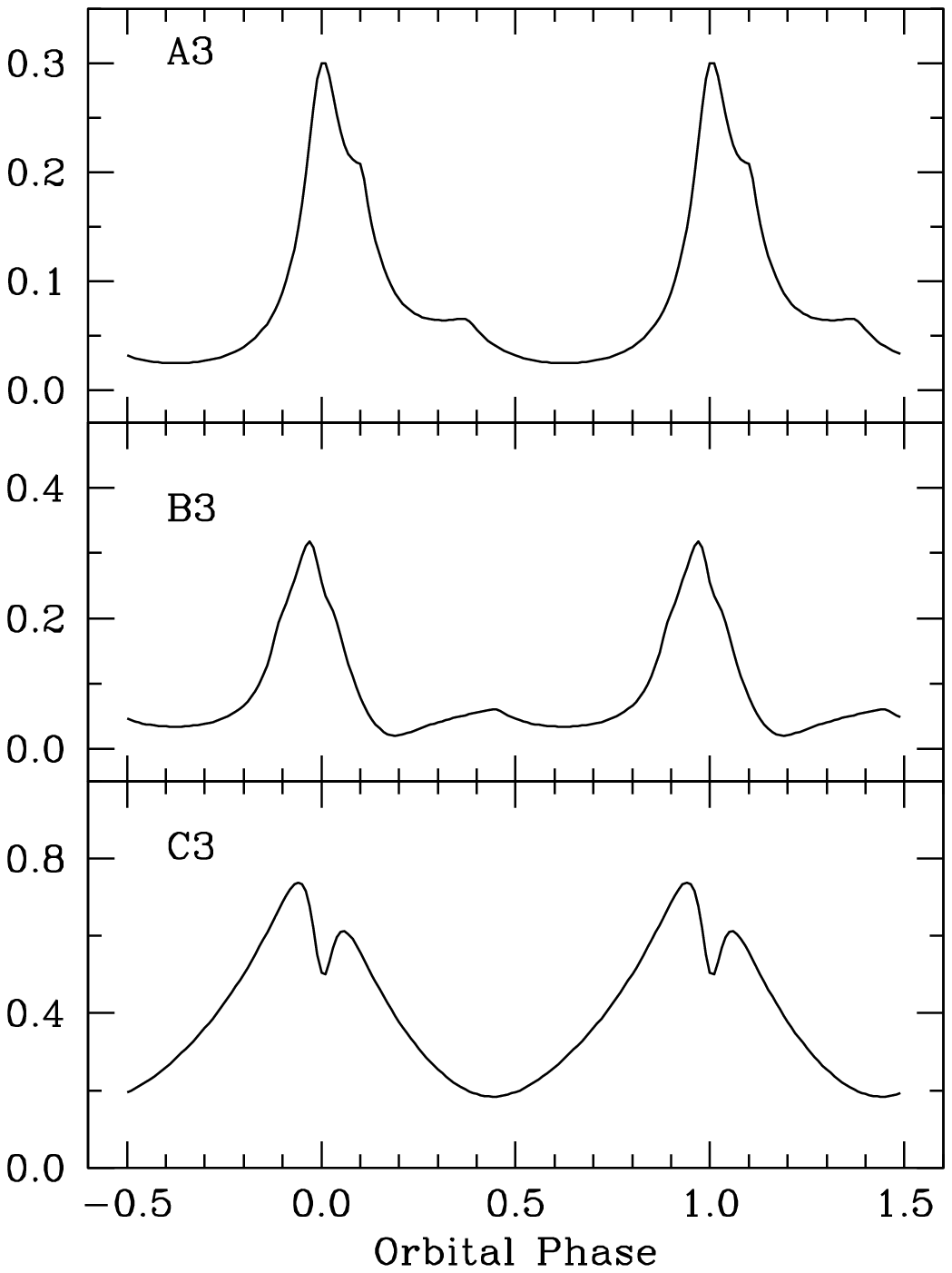}
 \caption[]{ Mass accretion rate using the Bondi-Hoyle formulae.
 Everywhere the orbital parameters are 
 $M_{1}=10\,M_{\odot}$, $M_{NS}=1.4\,M_{\odot}$,
 $P_{orb}$=26.4917 d, $e=0.6$. The wind parameters used and the calculated average mass accretion rate 
 are : \\
 {\bf (A1)} V$_0$=5 km s$^{-1}$, $n=3.25$, non-rotating disk, 
            $\overline {\dot M_a}=0.11\, \dot M_{loss}$\\ 
 {\bf (B1)} V$_0$=5 km s$^{-1}$, $n=3.25$, keplerian disk, 
            $\overline {\dot M_a}=2.9\,\dot M_{loss}$ \\ 
 {\bf (C1)} V$_0$=10 km s$^{-1}$, $n=2.1$, keplerian disk, 
            $\overline {\dot M_a}=9.7\, \dot M_{loss}$

The use of the Bondi-Hoyle formula gives  
obviously false values( at least in A1 and B1 ) values of the average mass accretion rate: 
 
{\bf (A2), (B2), (C2)} - the same parameters, but the factor of the filling of the accretion cylinder included.

{\bf (A3), (B3), (C3)} - the same parameters, but the upper limit set $R_a \leq r/4$, and the filling factor
included. The corresponding average mass accretion rates  are 
$\overline {\dot M_a} /\dot M_{loss}=$  0.09, 0.08, 0.36,  
for A3, B3, and C3 respectively.
}
  
 \label{Bondi}
\end{figure*}

As a result of the motion of the neutron star throw the Be circumstellar disk
the mass accretion rate will vary along the orbital period. A few examples,
calculated using different parameters for the Be circumstellar disk, are shown 
in Fig.\ref{Bondi}. 
During the calculations we assumed always an orbital eccentricity
$e=0.6$, a neutron star mass of $M_{NS}=1.4\,M_{\odot}$,
$M_{1}=10\,M_{\odot}$ and an orbital period $P_{orb}$=26.4917 days. 

As it is visible in Fig.\ref{Bondi}, we can obtain very strange mass accretion rates in
some cases. In Fig.\ref{Bondi}(b,c), the average mass accretion rate is
considerably higher than the mass loss rate of the Be star. This is a false result of 
the used formula. The Bondi-Hoyle description is valid for accretion from
infinite medium. It assumes that all the matter entering into a cylinder with radius
equal to the accretion radius ($R_a=\sqrt{2GM_{NS}/V_{rel}^2}$) will be accreted. 
In the context of a binary star, this is only valid provided that   
\begin{itemize}
      \item the motion is high supersonic $V_{rel}\,>>\,C_s$,
      \item $R_a << r$,
      \item the whole accretion cylinder is filled with  the outflowing wind.
\end{itemize}
 
Unfortunately these conditions are not always fulfilled in Be/X-ray binary systems. 

\subsection{ Mass accretion rate in Be/X-ray binaries }

As it was shown in the previous section the Bondi-Hoyle formulae are 
not giving good results in case of Be/X-ray binaries. The exact mass accretion 
rate could be found only if we consider simultaneously the formation of the wind
and the influence of the compact object onto the wind. 
Here we propose a modification of the mass accretion rate formula that is appropriate for
Be/X-ray binaries. In order to do so,  
the Bondi-Hoyle formula has to be rewritten in a way such that the
accretion radius explicitly appears: 
\begin{equation}
\dot M_a=\pi  R_a^2 \, \rho \, V_{rel}\,  ,
\end{equation}
where the density in an outflowing wind, with disk opening angle
$\theta$,  is given from the continuity equation: 
\begin{equation}
 \dot M_{loss}= 4 \pi r^2 \sin \theta \, V_w\, \rho, 
\end{equation}

The wind in the disk is most probably subsonic 
(i.e. Okazaki, 2001). A keplerian rotation in the disk will give in many cases
a low relative velocity between the NS and the wind.
This means that  the accretion radius depends also on the speed of sound, $c_s$ :   
\begin{equation}
 R_a  = \frac {2 G M_{NS}} {V_{rel}^2 + c_s^2}
\end{equation}

Using this formula (with or without $C_s^2$ term), 
the accretion radius of a neutron star can achieve values bigger than 
the distance between components. 
However the disks in the Be/X-ray binaries are more or less similar
to the disks in the "normal" Be stars (Zamanov et al. 2001) although some differences
exist. This means that the accretion radius can not be very big, otherwise 
the NS would control the motion of the wind and this is impossible
because the mass ratio is $M_1/M_{NS}\,>\,4$ (typically $M_1\,>\,8\,M_{\sun}$
and $M_{NS}=1.4 M_{\odot}$)   
We will thus adopt here an 
upper limit $R_a\,<\,0.25\,r$, where $r$ is the distance from the Be star   

Because the outflowing wind is confined in a disk, it is possible that the
vertical size of the disk is less than the accretion radius. Consequently, 
the  accretion cylinder would not entirely filled by the wind material. In case 
that $R_a\,> y$ where $y$ is the vertical size of the
outflowing disk $ y = r sin \theta $,
the accretion rate must be reduced with pure geometrical factor  
$ 2\pi R_a^{-2}[y \sqrt{R_a^2 - y^2} + R_a^2 \sin ^{-1} (y/R_a)]$.

We believe that the proposed modifications give considerably better results
concerning the mass accretion rate, i.e.
the NS will accrete a fraction of the wind. The results of the calculations
are demonstrated in Fig.\ref{Bondi} A3,B3,C3.

\section { X-ray variability in the light of Ejector-Propeller model}

  \begin{figure*}[htbp]
     \normalsize
     \vskip 3mm plus 1mm minus 1mm
           \epsfysize=9.0cm
           \centerline{\epsffile{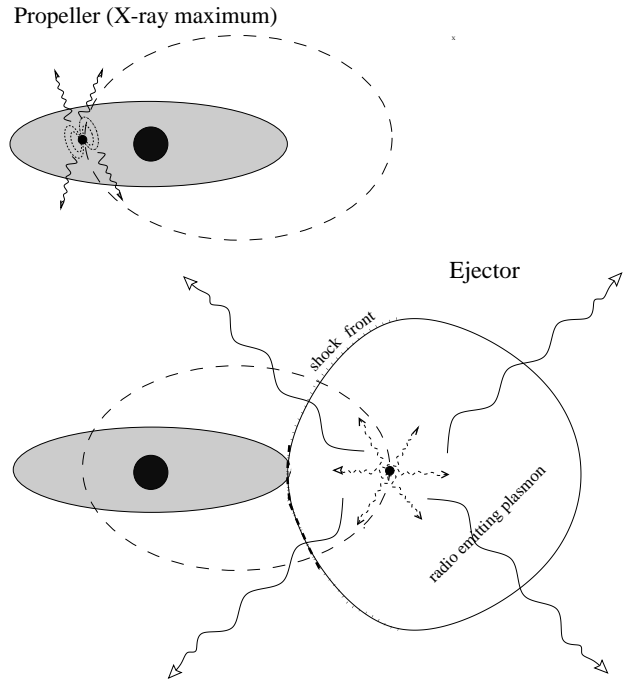}}
  \caption[] { A sketch of the Ejector-propeller model. The dashed lines indicates 
      the orbit of the neutron star.  Upper - 
      accretion onto the magnetosphere at periastron, 
      down - ejector and the cavern around the neutron star. 
              }    
  \label{Sketch}
  \end{figure*}
A sketch of the ejector-propeller model of \lsi\ is illustrated in Fig.\ref{Sketch}.
Following the suppositions the neutron star will 
act as propeller (accretion onto the magnetosphere) at periastron and as ejector
(young radio pulsar) at the apastron. 
The transitions between the regimes will depend how material is captured in the 
the neutron star 
accretion cylinder, possible formation of accretion disk during the accretion onto
the magnetosphere, and even the possible flip-flop instability of accretion
(Livio et al 1991). 

Following the assumptions of the ejector-propeller model, 
a lower mass capture rate is expected during the ejector regime.
In contrast, higher mass capture rates are expected during the
the propeller regime close to periastron passage.
During the propeller 
stage the X-ray luminosity can be estimated  as 
   \begin{equation}
     L_X (P)=2^{2/7}(GM)^{8/7}\mu^{-4/7}\dot M_a^{9/7},  
   \end{equation}
where G is the constant of gravitation, M is the mass of the neutron star, 
$\mu$ is its magnetic dipole moment (which is adopted 
$\mu=2\times 10^{30} $G\, cm$^3$), and $\dot M_a$ is the mass accretion rate. 

During the ejector stage efficiency 
of conversion of the spin-down energy into high energy emission can be 
expected to be about 1-10\% (Tavani et al. 1996). The estimated spin-down 
luminosity of the ejector is expected about $1-5\times 10^{35}$ erg\,s$^{-1}$,
so as reasonable value  can be adopted $L_X(E) = 2.10^{33} - 2.10^{34}$ erg\,s$^{-1}$.

The other possibility to estimate the X-ray luminosity of the expanding plasmon
is to consider Bremsstrahlung emission of the shocked gas.
In this case Taylor et al. (1996) also obtained  $2\times10^{34}$ erg\,s$^{-1}$.
The both estimates are in good agreement and are similar to the expected values 
of the magnetospheric accretion. 

It deserves to be noted that in case of transition between accretor and propeller 
a considerable luminosity gap will exist, about ~100 times difference in 
the X-ray luminosity, is expected and observed 
(Corbet 1996; Raguzova \& Lipunov, 1998). But this is not 
the case at the transition between ejector and propeller, where the X-ray luminosities 
are of the same order. So, we will assume that $L_X(ejector)=L_X(propeller)$ not 
to complicate our qualitative results.  

\begin{figure*}[htb]
 \mbox{}
 \vspace{15.0cm}
  \includegraphics{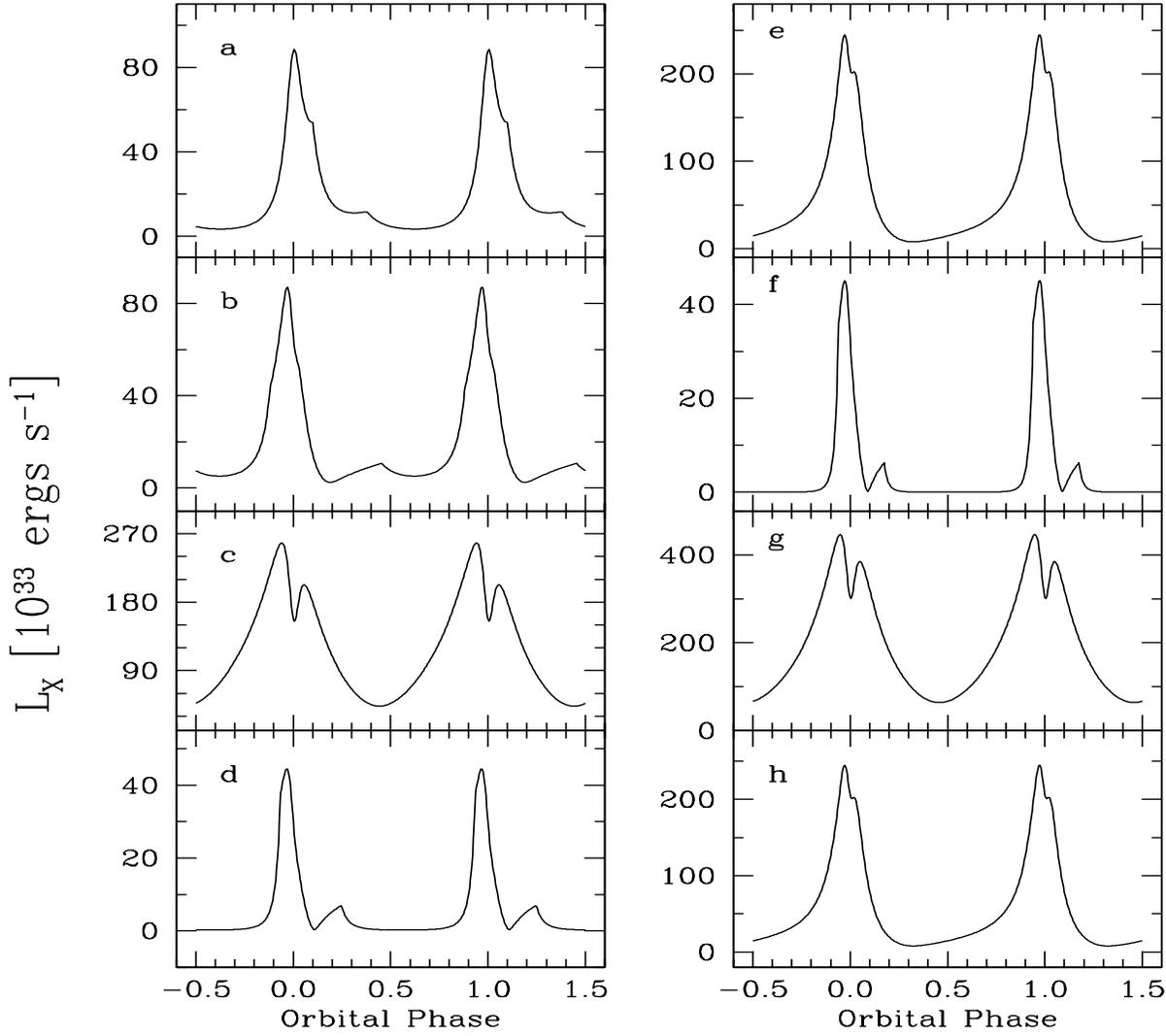}
 
\caption[]{The calculated X-ray luminosity of the the neutron star in \lsi\ 
           using different wind parameters.
           The $L_X$ is calculated as magnetospheric accretion luminosity and
	   supposing that $L_X(ejector) = L_X(propeller)$. The radio outbursts 
	   are expected to peak after the X-ray maximum, when the mass accretion rate is lower, 
	   and the NS can change its regime from propeller to ejector.  }
(a) V$_0$=5 km s$^{-1}$, $n=3.25$, non-rotating disk, 
                       $\overline {\dot M_a}=0.09\, \dot M_{loss}$\\ 
(b) V$_0$=5 km s$^{-1}$, $n=3.25$, keplerian disk, 
                       $\overline {\dot M_a}=0.08\, \dot M_{loss}$\\ 
(c) V$_0$=10 km s$^{-1}$, $n=2.1$, keplerian, 
                       $\overline {\dot M_a}=0.36\, \dot M_{loss}$\\ 
(d) V$_0$=10 km s$^{-1}$, $n=3.25$, keplerian, 
                       $\overline {\dot M_a}=0.03\, \dot M_{loss}$\\ 
(e) V$_0$=2 km s$^{-1}$, $n=3.25$, keplerian, 
                       $\overline {\dot M_a}=0.20\, \dot M_{loss}$\\ 
(f) V$_0$=5 km s$^{-1}$, $n=3.75$, keplerian, 
                       $\overline {\dot M_a}=0.02\, \dot M_{loss}$\\ 
(g) V$_0$=5 km s$^{-1}$, $n=2.25$, keplerian, 
                       $\overline {\dot M_a}=0.53\, \dot M_{loss}$\\ 
(h) V$_0$=2 km s$^{-1}$, $n=3.25$, keplerian, 
                       $\overline {\dot M_a}=0.20\, \dot M_{loss}$\\ 
Everywhere we adopted mass loss rate in the outflowing disk
$M_{loss}\,=10^{-9}\, M_{\odot}\,yr^{-1}$.

 \label{Xray}
\end{figure*}

The X-ray luminosity calculated in this way is plotted in Fig.\ref{Xray}. The orbital phase
zero corresponds to the periastron, the adopted mass loss rate of the Be star 
is $M_{loss}\,=10^{-9}\, M_{\odot}\,$ yr$^{-1}$. 
The calculated X-ray variability is very similar to the observed (Fig. \ref{Fiona}). 
The X-ray maximum is at periastron and this means that the periastron passage 
corresponds to radio phase $\sim 0.5$ using the modern radio ephemeris.

\section{Plasmon parameters and radio outbursts }

The switch on of the ejector
will create a cavern around the neutron star. This cavern will start to expand 
under the pressure of the relativistic wind of the ejector and can be identified 
with the expanding plasmon, which radio emission successfully fit
the observed radio outbursts of \lsi\ modeled by Paredes et al. (1991). 
The question here is: are the 
parameters expected at the propeller-ejector transition appropriate for the plasmon?

The most important parameters which appeared in the radio model of Paredes et al. (1991)
are initial radius ($R_0$), expansion velocity ($V_{\rm exp}$), initial magnetic field,
injection time interval, power law index of electrons, and the injection rate.
In the terms of Ejector-propeller model we can put constrains on the allowed ranges 
for this parameters:

- 1. The time of plasmon appearance - the plasmon will appear after the periastron 
passage at low mass accretion rate (for more details see Zamanov, 1995 and references
therein). So it means that the radio outbursts will peak always
with some phase shift after the change of the regime from propeller to ejector. 

- 2. The initial radius 
expected at the propeller-ejector transition 
will be of the order of light cylinder radius, and in any case
less than the accretion radius, which means that acceptable values are
of the order of $0.01-3.0\:R_\odot$ 

- 3. The expansion velocity of the cavern in terms of the
propeller-ejector transition can be estimated as (Zamanov, 1995):
    \begin{equation}
    V_{\rm exp} = V_{\rm w} + \sqrt{L_m V_w /(c \dot M_{loss})} 
    \end{equation}
where $c$ is the speed of light, $\dot M_{loss}$ and $V_w$ are the mass loss rate 
and the wind velocity of the Be star respectively, $L_m$ is magneto-dipole luminosity
of the young radio pulsar. Adopting $V_w=100\,$km\,s$^{-1}$ for the disk wind
and $V_w=1000\,$km\,s$^{-1}$ for the polar wind and
$L_m= 1-5\times 10^{35}$erg\,s$^{-1}$ the cavern expansion will be with velocity
$200-1300\,$km\,s$^{-1}$. 
In the radio light curves we will adopt values 
$V_{\rm exp}=$200, 400, 1200$\,$km\,s$^{-1}$. 

- 4. Injection time interval - it is the time neutron star acts as ejector. 
The switch on of the ejector will be when the neutron star goes away from the B star
and the switch off will be when it enters in the dense circumstellar disk
approaching the periastron. From the mass accretion rate behavior 
(Fig.\ref{Bondi} right  and Fig.\ref{Xray}) we can
expect injection time interval $0.4-0.7\times P_{orb}$ or in other words
10~$-$~20 days.

- 5. Injection rate - the total 
injection rate of the Crab into the nebula at a spin period of 
0.033 $s^{-1}$  is estimated to be about $10^{40}$-$10^{41}$ 
particles~$s^{-1}$ on base of the high-energy spectrum (De Jager \& Harding, 1992). 
From one side the spin period of the neutron star in \lsi\ 
required from the E-P model  is longer than Crab. 
We expect $P_{spin}=0.15-0.20\, s$ (see Zamanov 1995 for more details).
It means $\sim 100-1000$ times lower rate of  injected relativistic particles
($L_m \sim P_{spin}^{-4}$). 
From other side as a result of the magnetosphere accretion there 
are probably more particles for injection.  
We will adopt 5$\times$10$^{-38}\,e^-\,$s$^{-1}$

- 6. Magnetic field - the origin of magnetic field inside the plasmon will be 
something like to the magnetic field of the Crab nebula but some additional 
contribution from the B star is possible. The magnetic field referred in the 
"radio" figures (Fig.\ref{Josep1} ) are very high
because they refer to the magnetic field at the initial plasmon radius.
When the plasmon expands to sizes of 1 AU, as measured with VLBI, 
the magnetic field is then close to 1~Gauss in agreement with
VLBI estimates based on equipartition arguments.
(Note: Changes in the radius and the magnetic field do no 
alter the light curves provided that $B\,R^2$ is constant - magnetic flux
conservation).   

- 7. Electron power law index is adopted $p=1.6$,
in order to be consistent with observed non-thermal (i.e. negative) 
spectral indices when the plasmon becomes optically thin. 
According to synchrotron theory, the optically thin spectral index is
$\alpha=(1-p)/2$. 
In our case, we have then $\alpha=-0.3$ in agreement with typical observations. 

- 8. Energy range of relativistic electrons: \\ 
$m_e c^2\,<\,E\,< 10^{-2}\,erg$

- 9. Distance from the Earth: 2 kpc

Using the above parameters appropriate for the ejector-propeller model
and the plasmon prescription we generated radio light curves.
They are shown in Fig.\ref{Josep1}. The radio behavior
(amplitude and shape) is very similar
to the observed in the radio monitoring of \lsi.

\begin{figure*}[htb]
 \mbox{}
 \vspace{22.0cm}
  \includegraphics{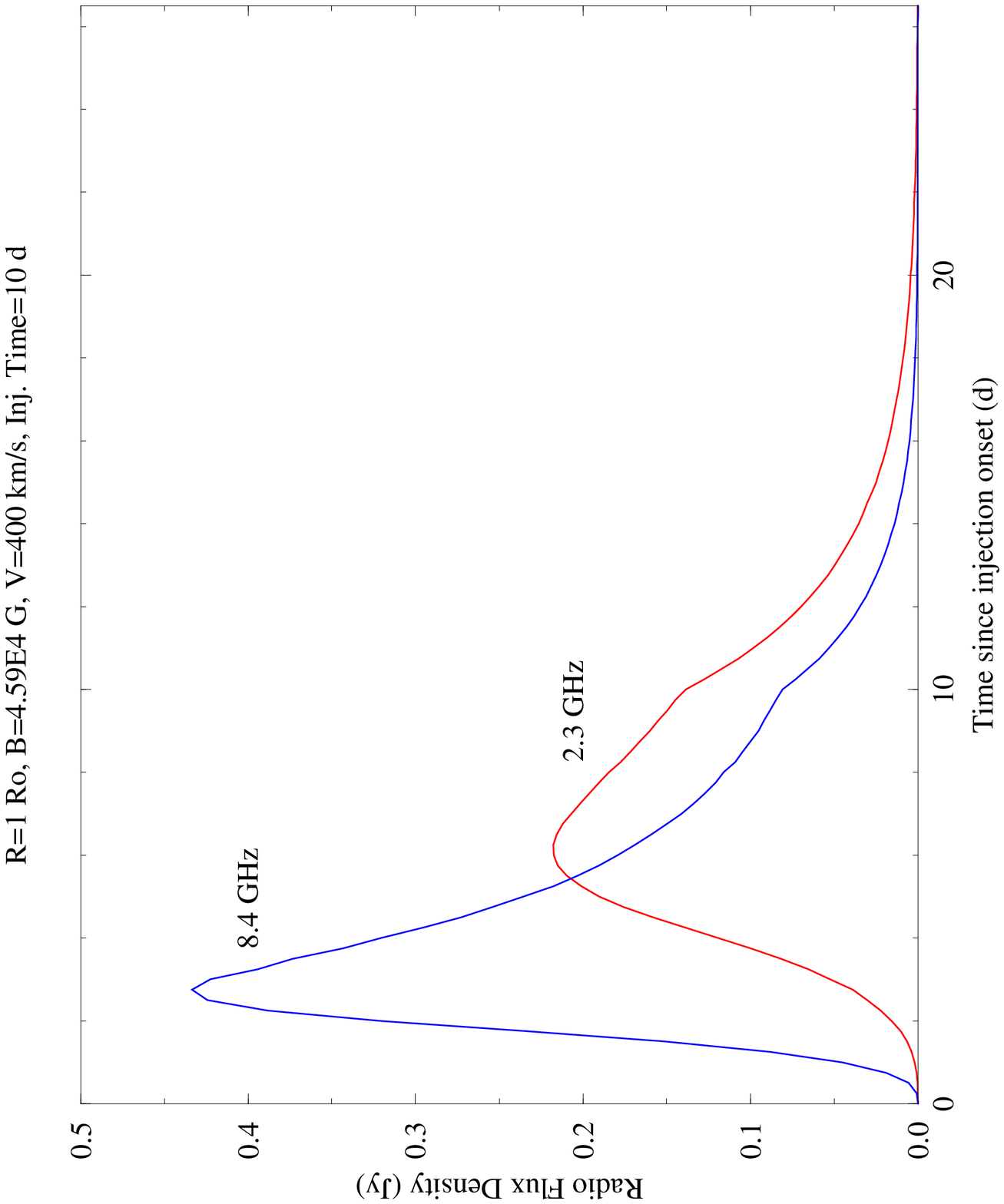}
  \includegraphics{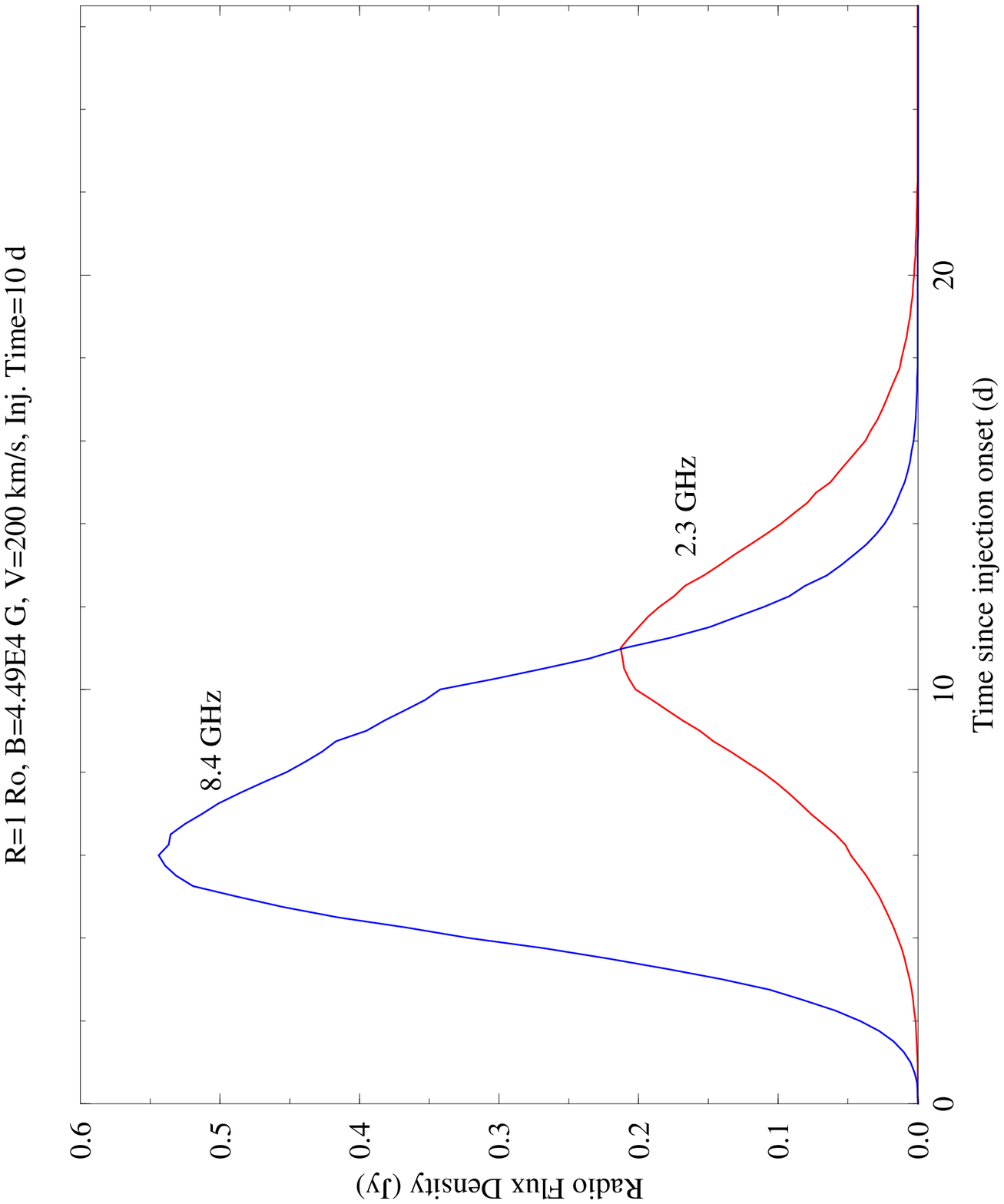} 
  \includegraphics{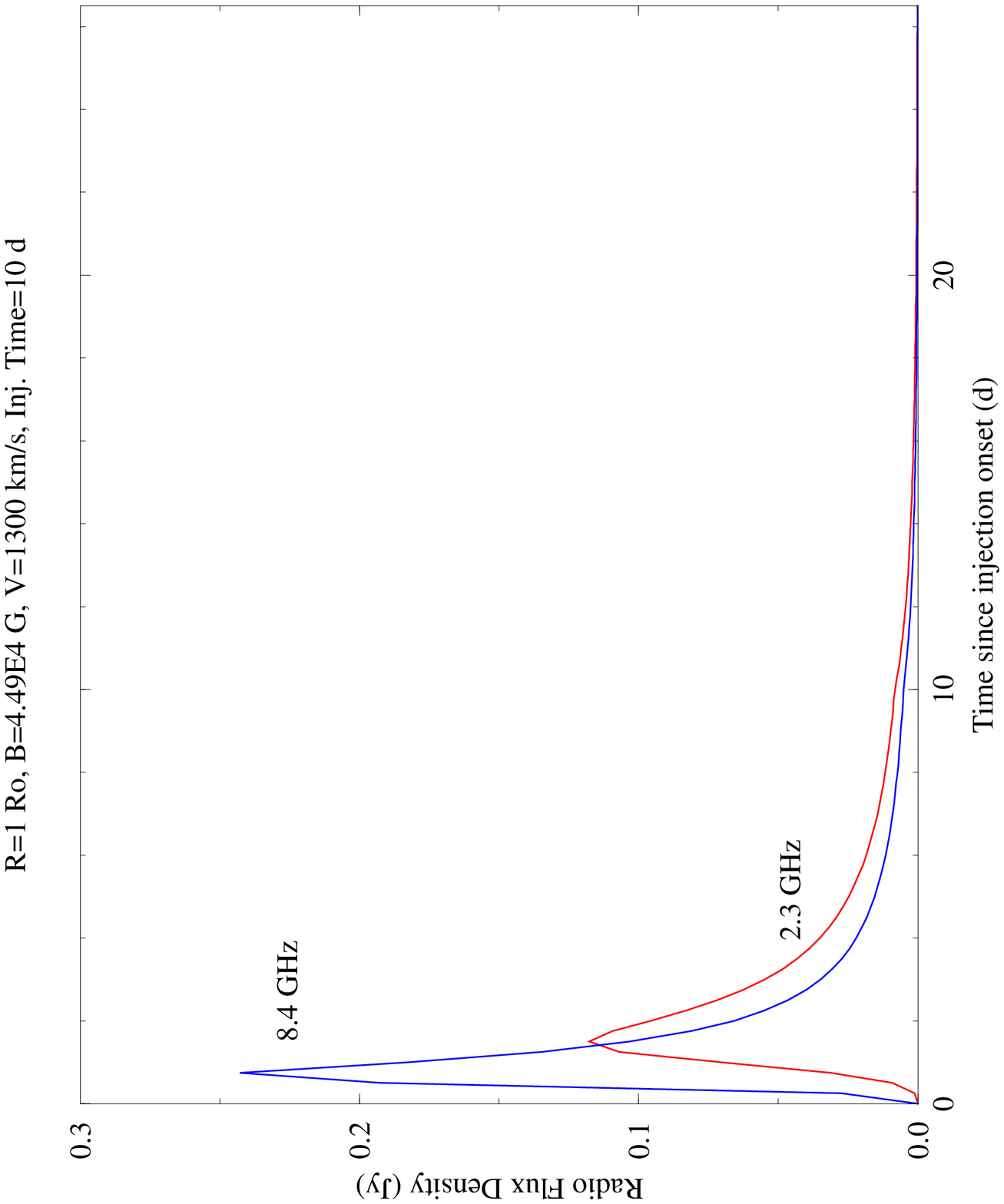}
  \includegraphics{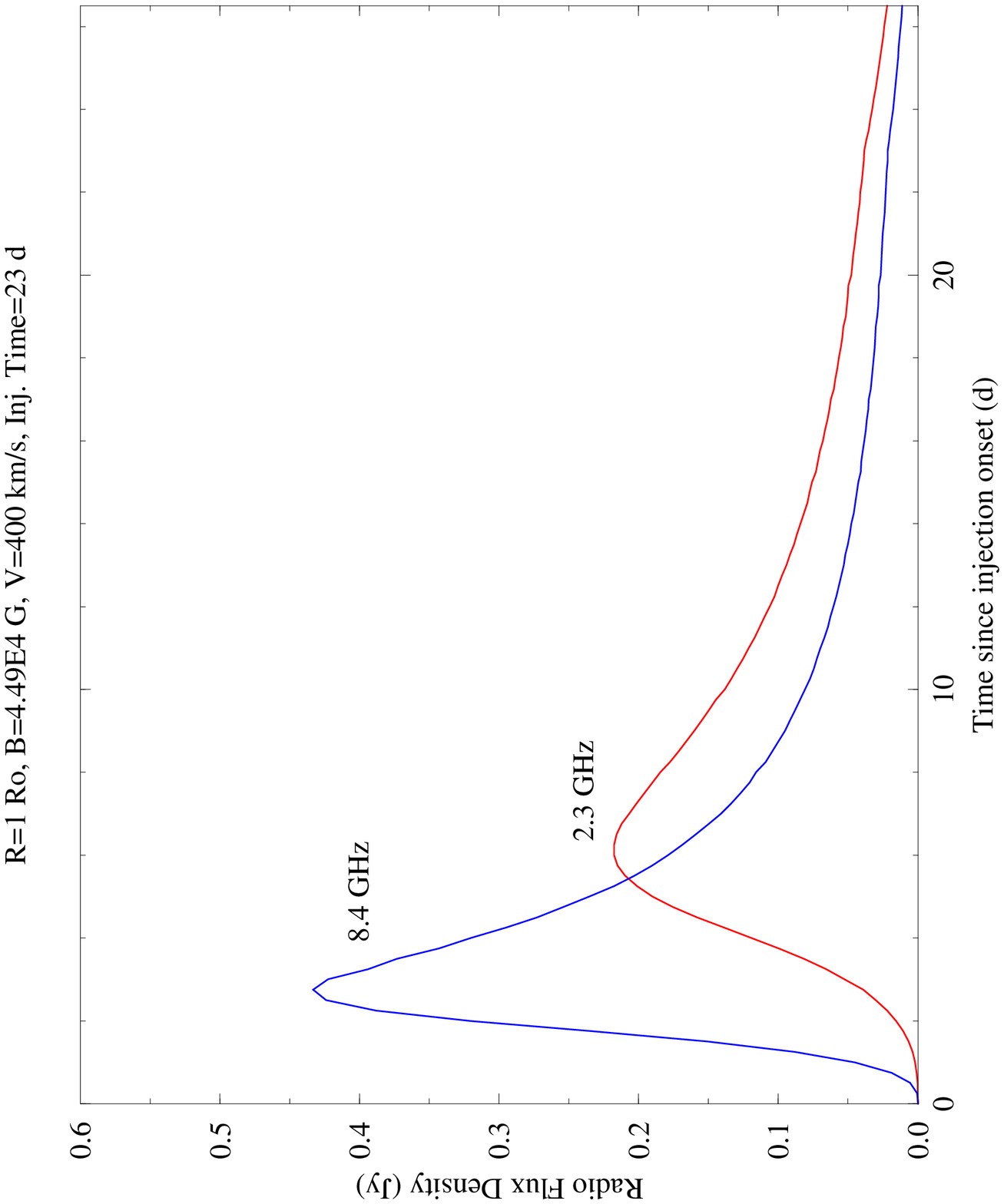} 

\caption[]{ Radio Light curves generated using the expanding plasmon
 and setting parameters appropriate for the ejector-propeller model. 
 }
 \label{Josep1}
\end{figure*}
\addtocounter{figure}{-1}
\begin{figure*}[htb]
 \mbox{}
 \vspace{22.0cm}
  \includegraphics{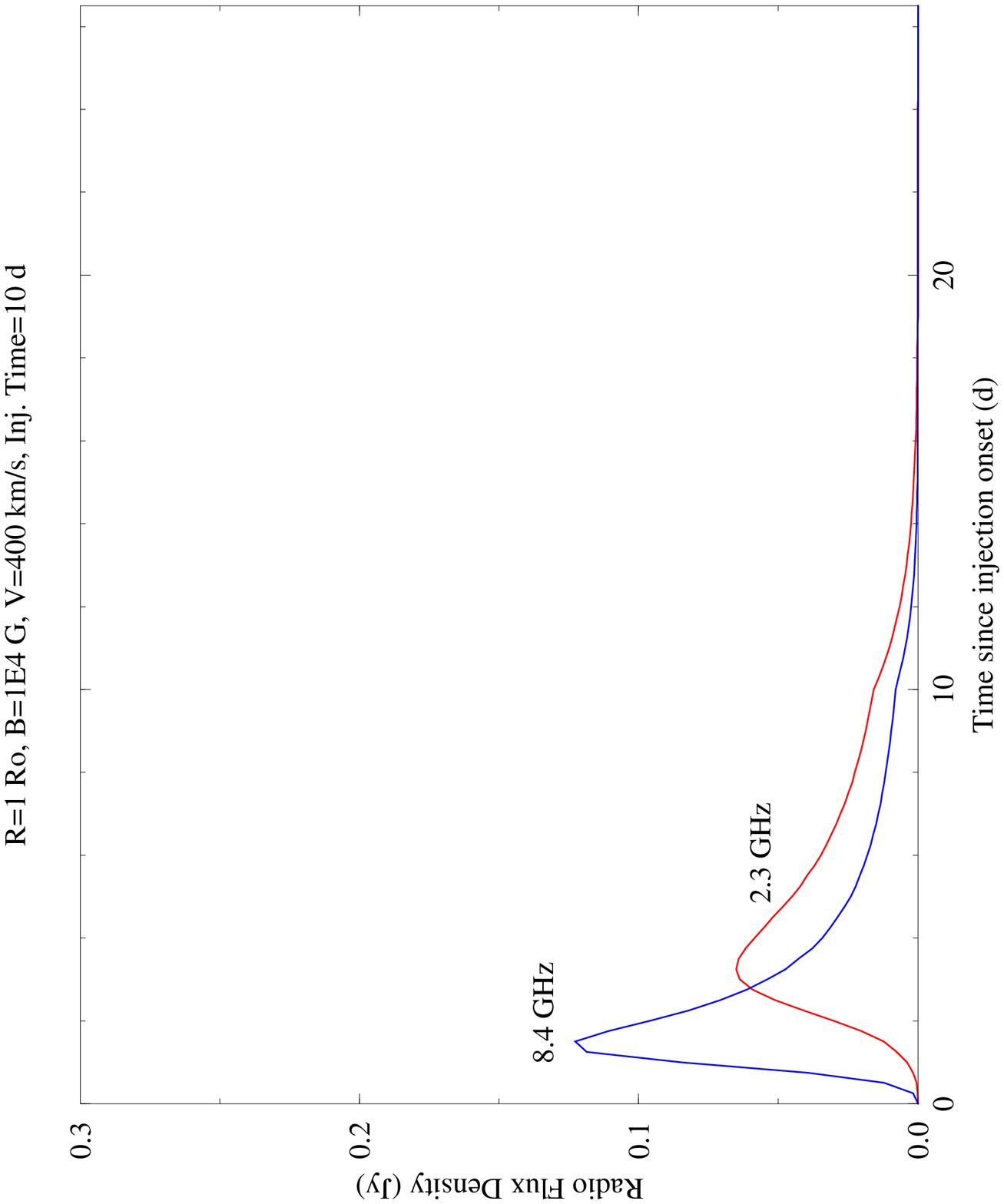}
  \includegraphics{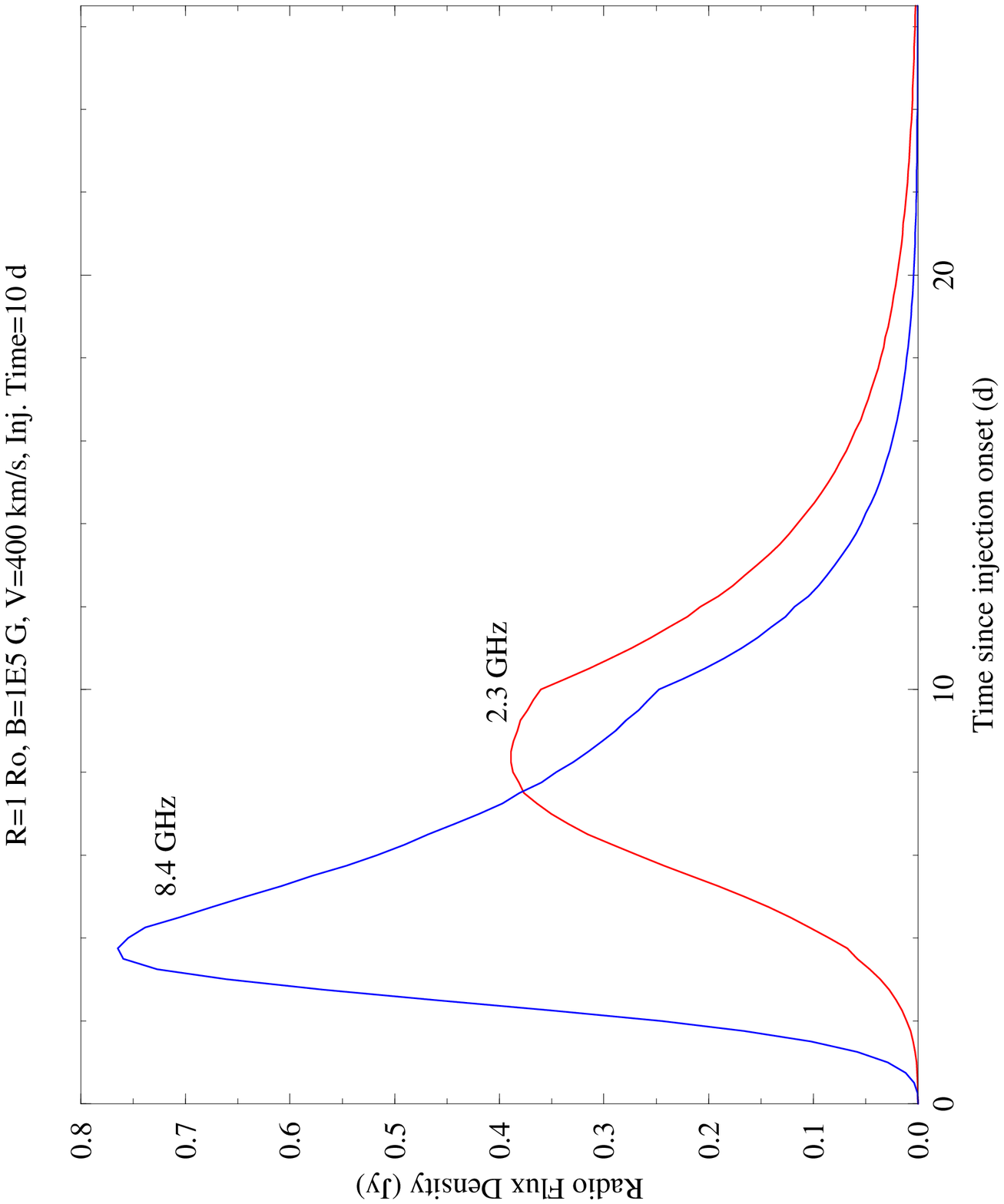} 
  \includegraphics{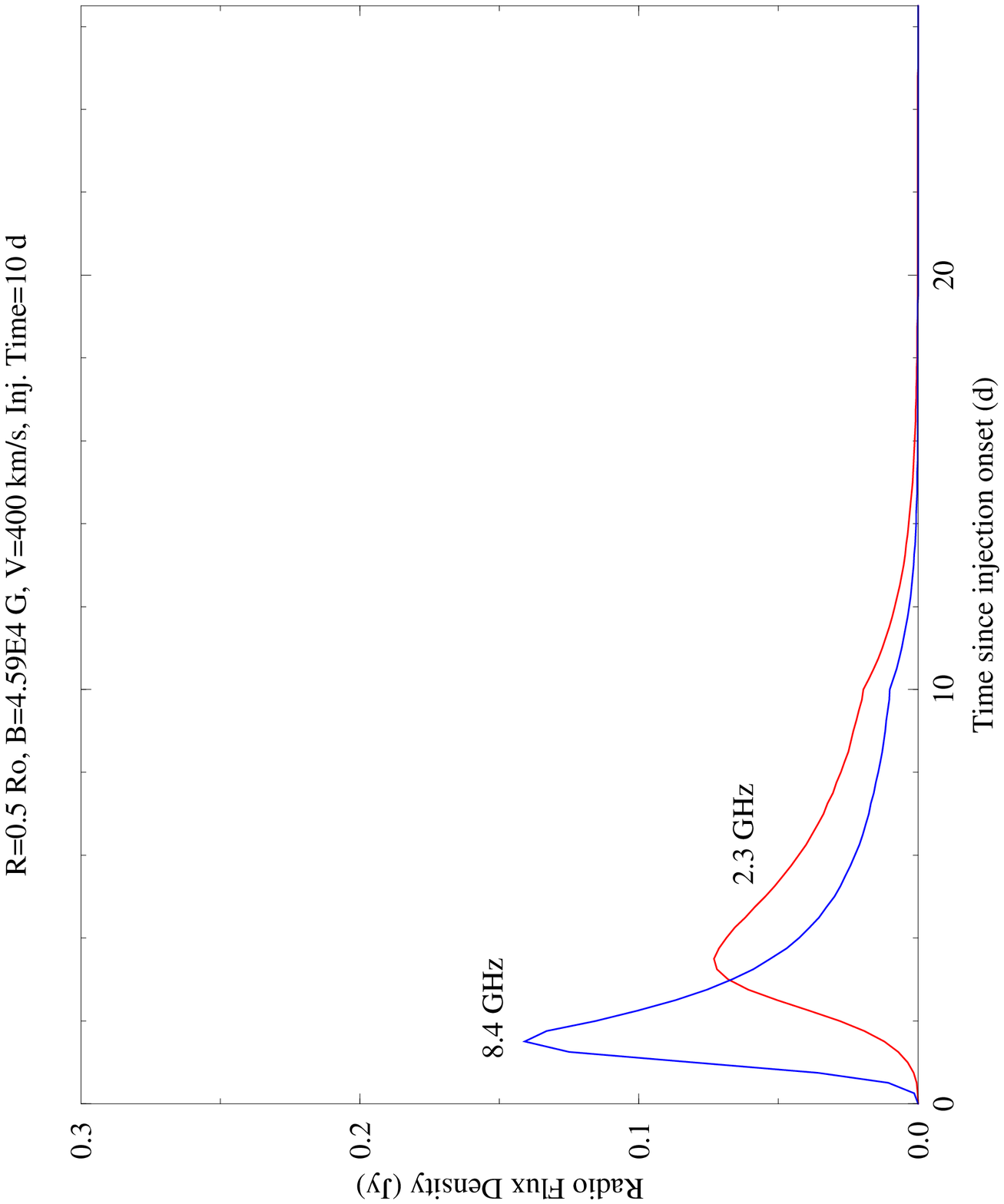}
  \includegraphics{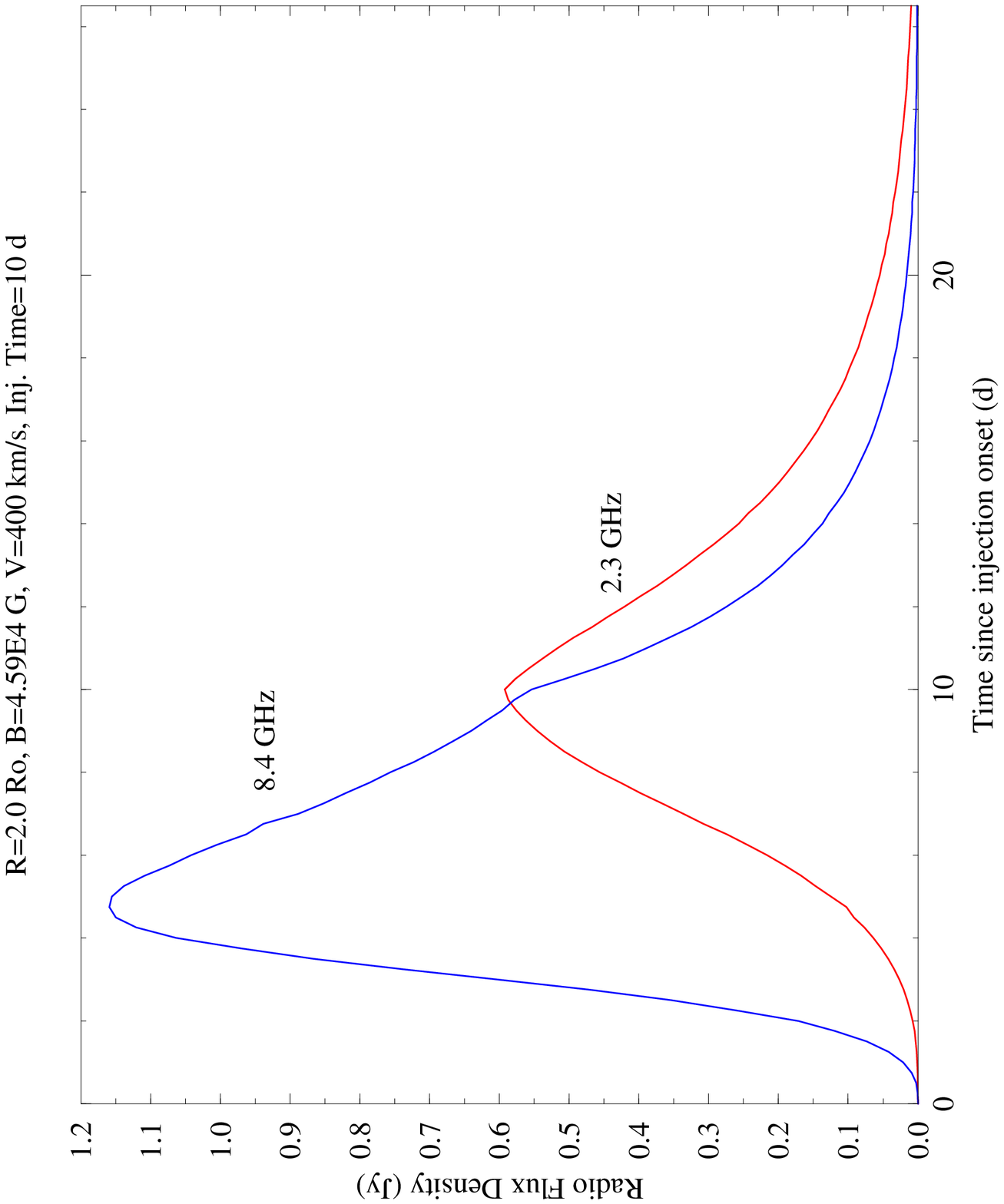} 

\caption[]{continued} 
 \label{Josep2}  
\end{figure*}

\section{Discussion}

Using the parameters appropriate to the ejector-propeller model 
we calculated the X-ray and radio variability of the Be/X-ray binary
\lsi. 
The general shape of the radio behavior (amplitude and shape) is very similar
to the observed in the radio monitoring of \lsi, as it was 
already shown in the "plasmon" calculations by Paredes et al.(1991). 
The radio outbursts peak can be achieved 2-8 days after the 
appearance of the plasmon (see Fig.\ref{Josep1} ). The appearance of the plasmon 
can be expected when the neutron star emerges from the denser part of the disk,
i.e. 2$-$5 days after the periastron (orbital period is 26.5 days).
In the calculated X-ray curves the X-ray maximum  corresponds to the 
periastron. The switch on of the ejector will be $0.1-0.4\,P_{orb}$ later.
It means that the radio peak will delay with 3-13 days after the X-ray maximum.
This is the behavior observed in both cases of the simultaneous 
radio and X-ray observations (Fig.\ref{Fiona}).
As can be seen in Fig.\ref{Xray}, the total X-ray luminosity  expected 
in  Ejector-Propeller model is of the order $1-200 \times 10^{33}\:$erg\,s$^{-1}$ 
This is in good agreement with the observed 
by ASCA  (Paredes et al. 1997) variability $1-6 \times 10^{34}\:$erg\,s$^{-1}$ 
in the 2-10 keV band. 
 
\subsection{High resolution radio maps}

Recent high resolution radio maps have evidenced 
a one sided radio jet at milliarcsecond scales (Massi et al. 2001).
These authors interpret it as a microquasar bipolar ejection with significant
Doppler boosting effect. The ejector-propeller model, explored
in this paper, provides however an alternative interpretation. 
Considering that the plasmon will be formed in one side of the Be star
(the apastron vicinity), a one sided radio jet is naturally expected.

\subsection{Asymmetry of the cavern}

To model the radio light curves we assumed that the plasmon 
expansion is spherically symmetric. But in the real case this region will be not
symmetric. Different forms of the cavern are possible - closed or open, 
for more details see Lipunov \& Prokhorov (1984). 
In addition, the fact that the neutron star ejects relativistic particles 
at the apastron of an eccentric orbit will lead to a radio source  
which is elongated in the direction of apastron. 
In any case, the radio source formed around the ejecting neutron star 
will be elongated in the direction controversial to the direction to the B star.

\subsection{Strong and weak radio outbursts}
It was discovered that radio outburst peak flux density varies over a time scale
of 1600 days. On the same time scale the $H\alpha$ emission 
of the outflowing disk (Zamanov \& Marti, 2000) varies as well,  and may be 
even the X-ray maximum (Apparao 2001).
Our preliminary tests evidenced
that a slower expansion velocity and stronger magnetic field 
give higher radio peak flux densities and the outburst peaks later.
Also a faster expansion velocity and lower magnetic field  
will result into lower peak flux densities and the earlier outburst peaks. 
These effects and their connection with the conditions in the Be disk
need of careful modeling.  

\subsection{ Multiple Outbursts }
A mechanism of multiple formation of caverns 
 around an ejecting neutron star is proposed (Lipunov \& Prokhorov, 1983).
We can speculate that, in case of multiple peaks in the \lsi\ radio outbursts,
we may be observing such a multiple cavern formation. Another possibility is a 
density structure in the wind of the Be star, i.e. rings with higher density
that can change the regime more than once during an orbital period. 

\subsection{Gamma rays} 
During the Ejector stage the gamma ray emission is thought to 
originate in the shock front at the boundary of the pulsar and stellar winds 
(Kniffen et al. 1997). Gamma ray production is possible during the 
propeller action too. Possible mechanisms are double layer formation in 
the interblob plasma (Wynn, King \& Horne, 1997) or high voltage at 
the boundaries of vortex structures (Wang and Robertson, 1985).  

\subsection{Other possibilities}

The fact that the radio outbursts are not exactly periodical point to they are a result
of a process which has appropriate conditions to begin somewhere at radio phases 
0.5-1.0, but when the process will switch on is a by-chance. We suppose that 
this process is the transition of the neutron star from propeller to ejector.  
Another alternative is the transition  closed-open cavern around the ejecting neutron star
(Zamanov 1995b, Harrison et al. 2000). In this case the expected magneto-dipole 
luminosity of the neutron star is about  $6 \times 10^{36}$~erg~s$^{-1}$, 
i.e. 10 times more than the expected in the ejector-propeller model.   

It can not be excluded that a magnetized black hole is acting in \lsi\ 
(Punsly, 1999). If such a compact object does exist in
\lsi, the microquasar scenario proposed by
Massi et al. (2001) should be also considered seriously. Further observations
are required in order to finally solve the true nature of this X-ray binary.

\section{Conclusions}

We tested the ejector-propeller model for the radio-emitting Be/X-ray binary
\lsi. The calculations show that the X-ray variability and 
radio outbursts can be the result of the transition from the
propeller to the ejector accretion regimes 
every orbital period. The main results are:

  - The parameters expected from the ejector-propeller model are appropriate
for the radio plasmon and the calculated radio light curves are similar to the observed
radio outbursts in \lsi. 

  - In terms of Ejector-Propeller model the X-ray maximum is a result of 
    the propeller action during the periastron passage of the neutron star.

  - The periastron is expected to correspond with radio phase $\sim 0.5$ using the 
    late radio ephemeris - i.e. the time of the X-ray maximum.
          
  - The observed phase shift between the radio and the X-ray maxima is in 
    agreement with the ejector-propeller model of \lsi. We expect that
    the X-ray and the radio maxima will always peak at different orbital phase.  

  - The total X-ray luminosity expected in terms of Ejector -Propeller model
    $1-200 \times 10^{33}\:$erg\,s$^{-1}$ is in good agreement with the observed 
    by ASCA  (Paredes et al. 1997) variability $1-6 \times 10^{34}\:$erg\,s$^{-1}$ 
    in the 2-10 keV band. 
  
There are a few systems where it is expected that the neutron star survives 
transitions from propeller to accretor and backward - A~0538-66, X~0331+53 
(Raguzova \& Lipunov, 1998). In our opinion, 
\lsi\ is probably the first example where we observe the transition from ejector 
to propeller and backward to occur every orbital period. 

During the lives of neutron stars in binary systems, it is believed that they 
evolve through different stages - ejector, propeller, accretor 
(Lipunov et al. 1994). 
Studies over systems such as \lsi, A~0538-66 and X~0331+53,  which probably 
exhibit transitions between these different regimes, will give us 
the unique possibility to investigate these transitions over a time scale of months.
This is an interval of time extremely short considering that the transitions, 
which are result of the neutron star evolution, occur 
on much longer time scales ($\sim10$ Myr). New light
over the evolution of the neutron stars in binary systems is expected from this kind
of work.

\begin{acknowledgements}

JM acknowledges partial support by  DGICYT (PB97-0903) and by Junta de Andaluc\'{\i}a (Spain).

\end{acknowledgements}

\end{document}